\documentclass[a4paper,11pt]{article}
\usepackage{pos}
\usepackage{amsmath}
\usepackage{braket}
\usepackage{array,multirow}
\usepackage{tikz}
\usepackage{dsfont}
\usetikzlibrary{snakes}
\usetikzlibrary{math}
\usetikzlibrary{calc}
\tikzset{decoration={snake,amplitude=.2mm,segment length=1mm,
		post length=0mm,pre length=0mm}}

\newcommand{\J}{{\tilde{J}}}
\newcommand{\f}[1]{\textbf{#1}}
\newcommand{\MatrixTwoXTwo}[4]{	\left( \begin{array}{cc} #1 & #2 \\ 
		#3 & #4 \\ \end{array}\right)}
\newcommand{\MatrixThreeXThree}[9]{	\left( \begin{array}{ccc} #1 & #2 & #3 \\ 
		#4 & #5 & #6 \\ 
		#7 & #8 & #9 \\\end{array}\right)}

\begin{document}
	
\title{Study of $I=0$ bottomonium bound states and resonances based on lattice QCD static potentials}
\ShortTitle{Study of $I=0$ bottomonium bound states and resonances}

\author[a]{Pedro Bicudo}
\author[a]{Nuno Cardoso}
\author*[b]{Lasse Mueller}
\author[b,c]{Marc Wagner}
\affiliation[a]{CeFEMA, Dep.\ F\'{\i}sica, Instituto Superior T\'ecnico, Universidade de Lisboa, \\ Av.\ Rovisco Pais, 1049-001 Lisboa, Portugal}

\affiliation[b]{Johann Wolfgang Goethe-Universit\"at Frankfurt am Main, Institut f\"ur Theoretische Physik, \\ Max-von-Laue-Stra{\ss}e 1, D-60438 Frankfurt am Main, Germany}

\affiliation[c]{Helmholtz Research Academy Hesse for FAIR, Campus Riedberg,  Max-von-Laue-Stra{\ss}e 12, \\
	D-60438 Frankfurt am Main, Germany}

\emailAdd{bicudo@tecnico.ulisboa.pt}
\emailAdd{nuno.cardoso@tecnico.ulisboa.pt}
\emailAdd{lmueller@itp.uni-frankfurt.de}
\emailAdd{mwagner@itp.uni-frankfurt.de}

\abstract{We investigate $I = 0$ bottomonium bound states and resonances in S, P, D and F waves using lattice QCD static-static-light-light potentials. We consider five coupled channels, one confined quarkonium and four open $B^{(*)}\bar{B}^{(*)}$ and $B^{(*)}_s\bar{B}^{(*)}_s$ meson-meson channels and use the Born-Oppenheimer approximation and the emergent wave method to compute poles of the \mbox{T} matrix. We discuss results for masses and decay widths and compare them to existing experimental results. Moreover, we determine the quarkonium and meson-meson composition of these states to clarify, whether they are ordinary quarkonium or should rather be interpreted as tetraquarks.}

\FullConference{
	The 39th International Symposium on Lattice Field Theory,\\
	8th-13th August, 2022,\\
	Rheinische Friedrich-Wilhelms-Universität Bonn, Bonn, Germany
}


\maketitle

\section{Introduction}

We discuss a comprehensive study of bottomonium bound states and resonances based on static-static-light-light potentials. We use the Born-Oppenheimer diabatic approximation \cite{Born:1927}, which is a two-step approach. First, static quark-antiquark potentials in presence of a light quark-antiquark pair are computed with lattice QCD (here we use existing results from Ref.\ \cite{Bali:2005fu}). Then, in a second step, these potentials are used in a coupled channel Schr\"odinger equation. This approach was successfully applied to $\bar{b}\bar{b}ud$ systems (see e.g.\ Refs.\ \cite{Bicudo:2016ooe,Bicudo:2017szl}) and to $I = 0$ bottomonium in an $S$ wave \cite{Bicudo:2019ymo,Bicudo:2020qhp}. There are also ongoing efforts to study $I = 1$ bottomonium in a similar way \cite{Prelovsek:2019ywc}.

In this contribution we extend our work \cite{Bicudo:2019ymo,Bicudo:2020qhp} from $S$ wave to $P$, $D$ and $F$ wave states. More details can be found in Ref.\ \cite{Bicudo:2022ihz}. See also Refs.\ \cite{Bruschini:2021sjh,Bruschini:2021ckr,TarrusCastella:2022rxb} for similar works by other independent groups.

\section{Coupled channel Schr\"odinger equation}

	In the following we briefly discuss the coupled channel Schr\"odinger equation for $I = 0$ bottomonium bound states and resonances. We consider a quarkonium channel $\bar{Q}Q$, a heavy-light meson pair $\bar{M}M$ with $u/d$ light quarks, isospin $I = 0$ (i.e.\ $\bar{Q}Q(\bar{u}u+\bar{d}d)$) and a heavy-light meson pair $\bar{M}_s M_s$ with $s$ light quarks (i.e.\ $\bar{Q}Q\bar{s}s$).

	Throughout this work we use the following quantum numbers:
\begin{itemize}
	\item $J^{PC}$: total angular momentum, parity and charge conjugation.
	\item $S^{PC}_Q$: spin of $\bar{Q}Q$ and corresponding parity and charge conjugation.
	\item $\J^{PC}$: total angular momentum excluding the heavy $\bar{Q}Q$ spins and corresponding parity and charge conjugation (for quarkonium $\J^{PC}$ coincides with the orbital angular momentum $L^{PC}$ of the two heavy quarks).
\end{itemize}
Since we treat the heavy quark spins as conserved quantities, energy levels and other observables do not depend on $S^{PC}_Q$. Thus, the relevant quantum numbers in our work to label bottomonium are $\J^{PC}$, not $J^{PC}$ as usual.

The Schr\"odinger equation has a 7-component wave function \\ $\psi(\mathbf{r}) = (\psi_{\bar{Q}Q}(\mathbf{r}), \vec{\psi}_{\bar{M}M}(\mathbf{r}), \vec{\psi}_{\bar{M_s}M_s}(\mathbf{r}))$. The first component represents the $\bar Q Q$-channel, while the six components below represent the respective $\bar M M$ and $\bar M _s M_s $ triplets with light spin $1$. The Schr\"odinger equation reads
\begin{align}
\left(-\frac{1}{2} \mu^{-1} \bigg(\partial_r^2 + \frac{2}{r} \partial_r - \frac{\f{L}^2}{r^2}\bigg) + V(\f{r}) + \MatrixThreeXThree{E_{\text{threshold}}}{0}{0}{0}{2m_M}{0}{0}{0}{2m_{M_s}}-E\right) \psi(\f{r}) = 0 , \label{eqn:Schr\"odinger_equation}
\end{align}
where $\mu^{-1} = \text{diag}(1/\mu_Q, 1/\mu_M, 1/\mu_M, 1/\mu_M, 1/\mu_{M_s}, 1/\mu_{M_s}, 1/\mu_{M_s})$ is a $7 \times 7$ diagonal matrix with the reduced masses, corresponding to a heavy quark-antiquark pair and to meson-meson pairs, i.e.\ $\mu_Q = m_Q/2$, $\mu_M = m_M/2$ and $\mu_{M_s} = m_{M_s}/2$ (we use spin averaged masses for $m_M$ and $m_{M_s}$). $\f{L} = \f{r} \times \f{p}$ denotes the orbital angular momentum operator and $E_{\text{threshold}}$ is the threshold energy corresponding to two negative parity static-light mesons in the same lattice setup, where also the static potentials were computed (for details see Ref.\ \cite{Bicudo:2020qhp}). The $7 \times 7$ potential matrix $V(\f{r})$ is given by
\begin{align}
\label{EQN001} V(\f{r}) = \MatrixThreeXThree{V_{\bar{Q}Q}(r)}{V_{\text{mix}}(r)\left(1\otimes\f{e}_r\right)}{(1/\sqrt{2}) V_{\text{mix}}(r)\left(1\otimes\f{e}_r\right)}
{V_{\text{mix}}(r)\left(\f{e}_r\otimes 1\right)}{V_{\bar{M}M}(r)}{0}
{(1/\sqrt{2}) V_{\text{mix}}(r)\left(\f{e}_r\otimes 1\right)}{0}{V_{\bar{M}M}(r)}
\end{align}
with
\begin{align}
\label{EQN002} V_{\bar{M}M}(r) = V_{\bar{M}M, \parallel}(r) (\f{e}_r\otimes\f{e}_r) + V_{\bar{M}M, \perp}(r) (1-\f{e}_r\otimes\f{e}_r) .
\end{align}
$V_{\bar{Q}Q}$, $V_{\text{mix}}$, $V_{\bar{M}M, \parallel}$ and $V_{\bar{M}M, \perp}$ can be computed with lattice QCD (see section~\ref{SEC001}).

	\section{\label{SEC001}Static potentials from lattice QCD}
	\newcommand{\tikzscale}{0.4}

	As input for our work we use static potentials computed with lattice QCD in the context of string breaking in Ref.\ \cite{Bali:2005fu}. The basic principle to compute such potentials is to define suitable creation operators for a quark-antiquark pair and a meson-meson pair, e.g.\
	\begin{eqnarray}
		& & \hspace{-0.7cm} \mathcal{O}_{Q\bar{Q}} = (\Gamma_Q)_{AB} \left(\bar{Q}_{A}(\mathbf{0})\;U(\mathbf{0}; \mathbf{r})\;Q_B(\mathbf{r})\right) \\
		& & \hspace{-0.7cm} \mathcal{O}_{M\bar{M}} = (\Gamma_Q)_{AB}(\Gamma_q)_{CD} \left(\bar{Q}_{A}(\mathbf{0})\;u_D (\mathbf{0})\;\bar{u}_C(\mathbf{r})\; Q_B(\mathbf{r})\right) + (u \rightarrow d) ,
	\end{eqnarray}
	and to compute the corresponding correlation matrix
	\begin{align}
		C(t) &= \MatrixTwoXTwo{\braket{\mathcal{O}_{Q\bar{Q}}|\mathcal{O}_{Q\bar{Q}}}_U}
		{\braket{\mathcal{O}_{Q\bar{Q}}|\mathcal{O}_{M\bar{M}}}_U}
		{\braket{\mathcal{O}_{M\bar{M}}|\mathcal{O}_{Q\bar{Q}}}_U}
 		{\braket{\mathcal{O}_{M\bar{M}}|\mathcal{O}_{M\bar{M}}}_U}
		= \MatrixTwoXTwo
		{
			\begin{tikzpicture}[scale = \tikzscale]
			\draw (0,0) -- (1,0) -- (1,1.3) -- (0,1.3) -- (0,0);
			\end{tikzpicture}
		}
		{	
			\begin{tikzpicture}[scale = \tikzscale]
			\draw[white] (0,0) rectangle (0.5,1.3);
			\node at (0.25,0.65) {$\sqrt{2}$};
			\end{tikzpicture}
			\begin{tikzpicture}[scale = \tikzscale]
			\draw (0,0) -- (0,1.3);
			\draw[decorate] (0,1.3) -- (1,1.3);
			\draw (1,1.3) -- (1,0);
			\draw (1,0) -- (0,0);
			\end{tikzpicture}	
		}
		{	
			\begin{tikzpicture}[scale = \tikzscale]
			\draw[white] (0,0) rectangle (0.5,1.3);
			\node at (0.25,0.65) {$\sqrt{2}$};
			\end{tikzpicture}
			\begin{tikzpicture}[scale = \tikzscale]
			\draw (0,0) -- (0,1.3);
			\draw (0,1.3) -- (1,1.3);
			\draw (1,1.3) -- (1,0);
			\draw[decorate] (1,0) -- (0,0);
			\end{tikzpicture}
		}
		{	
			\begin{tikzpicture}[scale = \tikzscale]
			\draw[white] (0,0) rectangle (0.5,1.3);
			\node at (0.25,0.65) {-$2$};
			\end{tikzpicture}
			\begin{tikzpicture}[scale = \tikzscale]
			\draw (0,0) -- (0,1.3);
			\draw[decorate] (0,1.3) -- (1,1.3);
			\draw (1,1.3) -- (1,0);
			\draw[decorate] (1,0) -- (0,0);
			\end{tikzpicture}
			\begin{tikzpicture}[scale = \tikzscale]
			\draw[white] (0,0) rectangle (0.5,1.3);
			\draw (0.1,0.65) -- (0.4,0.65);
			\draw (0.25,0.5) -- (0.25,0.8);
			\end{tikzpicture}
			\begin{tikzpicture}[scale = \tikzscale]
			\draw (0,0) -- (0,1.3);
			\draw[decorate] (0,1.3) to [bend left=45] (0,0);
			\draw (1,1.3) -- (1,0);
			\draw[decorate] (1,0) to [bend left=45] (1,1.3);
			\end{tikzpicture}
		}.
	\label{eqn:correlation_matrix}
	\end{align}
	Solid lines after the last equality sign in Eq.\ \eqref{eqn:correlation_matrix} indicate gluonic parallel transporters, while wiggly lines correspond to light quark propagators. Thus, the upper left matrix element is a Wilson loop, the off-diagonal matrix elements are similar to Wilson loops with one gluonic parallel transporter replaced by a light quark propagator and the lower right matrix element is a sum of two fermionic diagrams, one connected and the other disconnected.
	From $C(t)$ the ground state potential $V_0(r)$ and the first excitation $V_1(r)$ can be extracted in the limit of large temporal separations using the spectral decomposition
	\begin{align}
		C_{jk}(t) = \sum_{n} a_{jk}^n(r)\text{e}^{-V_n(r)t} .
	\end{align}
	
\begin{figure}
		\centering
		\includegraphics[width=0.7\textwidth]{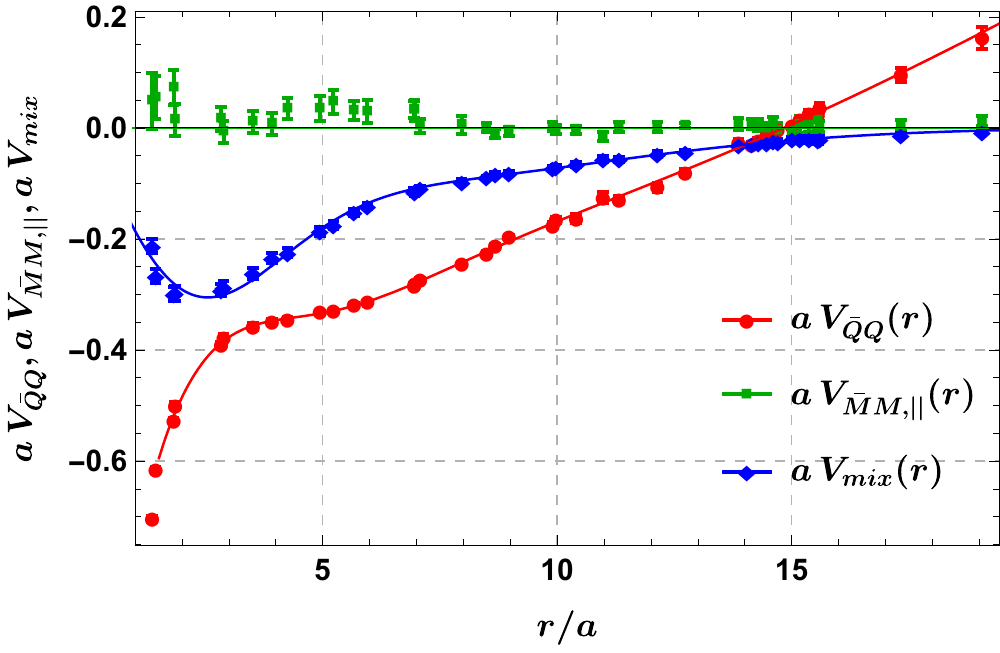}
		\caption{Potentials $V_{\bar{Q}Q}$, $V_{\bar{M}M, \parallel}$ and $V_{\text{mix}}$ as functions of the $\bar Q Q$ separation $r$. The curves represent the parameterizations \eqref{eqn:parameterization1} to \eqref{eqn:parameterization3} with parameters listed in Table \ref{tab:fitsGevFm}.}
		\label{fig:potentials}
	\end{figure}

  The relation between $V_0(r)$ and $V_1(r)$ and the potentials appearing in Eqs.\ (\ref{EQN001}) and (\ref{EQN002}) was derived in Ref.\ \cite{Bicudo:2019ymo} and is given by
	\begin{eqnarray}
		\hspace{-0.7cm} & & V_{\bar{Q}Q}(r) =\cos^2(\theta(r))V_0(r) + 
		\sin^2(\theta(r))V_1(r) \\
		\hspace{-0.7cm} & & V_{\bar{M}M, \parallel}(r) =\sin^2(\theta(r))V_0(r) + 
		\cos^2(\theta(r))V_1(r) \\
		\hspace{-0.7cm} & & V_{\text{mix}}(r) =\cos(\theta(r))\sin(\theta(r))\Big(V_0(r) + 
		V_1(r)\Big) \\
		\hspace{-0.7cm} & & V_{\bar{M}M, \perp}(r) = 0.
	\end{eqnarray}			
	In Fig.\ \ref{fig:potentials} we show the lattice data points for $V_{\bar{Q}Q}(r)$, $V_{\text{mix}}(r)$ and $V_{\bar{M}M}(r)$ and parameterizations
	\begin{eqnarray}
		& & \hspace{-0.7cm} V_{\bar{Q}Q}(r) = E_0 - \frac{\alpha}{r} + \sigma r + \sum_{j=1}^{2} c_{\bar{Q}Q, j} \, r \exp\bigg(-\frac{r^2}{2\lambda^2_{\bar{Q}Q, j}}\bigg) \label{eqn:parameterization1}\\
		& & \hspace{-0.7cm} V_{\bar{M}M, \parallel}(r) = 0 \\	
		& & \hspace{-0.7cm} V_{\text{mix}}(r) = \sum_{j=1}^{2} c_{\text{mix}, j} \, r \exp\bigg(-\frac{r^2}{2\lambda^2_{\text{mix}, j}}\bigg)
		\label{eqn:parameterization3}
	\end{eqnarray}
	with parameters listed in Table \ref{tab:fitsGevFm}.

	\begin{table}[htb]
		\centering
			\begin{tabular}{c|c|c}
				\hline
				potential & parameter & value \\ 
				\hline
				\hline
				$V_{\bar{Q} Q}(r)$ & $E_0$                      & $-1.599(269) \, \text{GeV}\phantom{1.^{-1}}$ \\
				& $\alpha$                   & $+0.320(94) \phantom{1.0 \, \text{GeV}^{-1}}$ \\
				& $\sigma$                   & $+0.253(035) \, \text{GeV}^{2\phantom{-}}\phantom{1.}$ \\
				& $c_{\bar{Q} Q,1}$          & $+0.826(882) \, \text{GeV}^{2\phantom{-}}\phantom{1.}$ \\
				& $\lambda_{\bar{Q} Q,1}$    & $+0.964(47) \, \text{GeV}^{-1}\phantom{1.0}$ \\
				& $c_{\bar{Q} Q,2}$          & $+0.174(1.004) \, \text{GeV}^{2\phantom{-}}$ \\
				\hline
				& $\lambda_{\bar{Q} Q,2}$    & $+2.663(425) \, \text{GeV}^{-1}\phantom{1.}$ \\
				\hline
				\hline
				$V_{\bar{M} M,\parallel}(r)$ & -- & -- \\
				\hline
				\hline
				$V_{\text{mix}}(r)$ & $c_{\text{mix},1}$       & $-0.988(32) \, \text{GeV}^{2\phantom{-}}\phantom{1.0}$ \\
				& $\lambda_{\text{mix},1}$ & $+0.982(18) \, \text{GeV}^{-1}\phantom{1.0}$ \\
				& $c_{\text{mix},2}$       & $-0.142(7) \, \text{GeV}^{2\phantom{-}}\phantom{1.00}$ \\
				\hline
				& $\lambda_{\text{mix},2}$ & $+2.666(46) \, \text{GeV}^{-1}\phantom{1.0}$ \\
			\end{tabular}
		\caption{\label{tab:fitsGevFm}Parameters of the potential parametrizations (\ref{eqn:parameterization1}) to (\ref{eqn:parameterization3}).}
	\end{table}

\section{Schr\"odinger equation and $\mbox{T}$ matrix for definite $\J$}

	By expanding $\psi(\mathbf{r})$ in terms of eigenfunctions of the operator corresponding to $\J$ one can project the Schr\"odinger equation \eqref{eqn:Schr\"odinger_equation} to definite $\J$ (for details see Refs.\ \cite{Bicudo:2019ymo,Bicudo:2022ihz}). This leads to a set of five coupled ordinary differential equations in the radial coordinate $r$,
	\begin{eqnarray}
		\nonumber & & \hspace{-0.7cm} \bigg(\frac{1}{2}\mu^{-1}\,\partial_r^2 + \frac{1}{2r^2}L_{\J}^2 + V_{\J}(r) + \\
		\nonumber & & \hspace{0.675cm} +\left.
		\left(\begin{array}{ccccc}
		E_{\text{threshold}} & 0 & 0 & 0 & 0 \\
		0 & 2m_M & 0 & 0 & 0 \\
		0 & 0 & 2m_M & 0 & 0 \\
		0 & 0 & 0 & 2m_{M_s} & 0 \\
		0 & 0 & 0 & 0 & 2m_{M_s} \\
		\end{array}\right) 
		- E\;\mathds{1}_{5\times 5} \right)
		\left( \begin{array}{c} u_{\J}(r) \\ \chi_{\bar{M}M,\J-1 \rightarrow \J}(r) \\ \chi_{\bar{M}M,\J+1 \rightarrow \J}(r) \\ \chi_{\bar{M_s}M_s,\J-1 \rightarrow \J}(r) \\ \chi_{\bar{M_s}M_s,\J+1 \rightarrow \J}(r) \end{array}\right) = \\
		\nonumber & & = \left( \begin{array}{c} V_{\text{mix}}(r) \\ 0 \\ 0 \\ 0 \\ 0 \end{array}\right) \bigg( \alpha_{\bar{M}M,1} \; {\J \over 2 \J+1} \; r \; j_{\J-1}(kr) + \alpha_{\bar{M}M,2} \; {\J+1 \over 2\J+1} \; r \; j_{\J+1}(kr) + \\
		 & & \hspace{0.675cm} + \alpha_{\bar{M_s}M_s,1} \; {\J \over 2 \J+1} \frac{r \; j_{\J-1}(k_sr)}{\sqrt{2}} + \alpha_{\bar{M_s}M_s,2} \; {\J+1 \over 2\J+1} \frac{r \; j_{\J+1}(k_sr)}{\sqrt{2}}\bigg) ,
		\label{eqn:coupledChannelSE_5x5}
	\end{eqnarray}
	where 
	\begin{eqnarray}
		\hspace{-0.7cm} & & \mu^{-1} = \text{diag}(1/\mu_Q, 1/\mu_M, 1/\mu_M, 1/\mu_{M_s}, 1/\mu_{M_s}) \\
		\hspace{-0.7cm} & & L_\J^2 = \text{diag}(\J(\J+1), (\J-1)\J, (\J+1)(\J+2), (\J-1)\J, (\J+1)(\J+2))
	\end{eqnarray}
	and
	\begin{eqnarray}
		\nonumber & & \hspace{-0.7cm} V_\J(r) = \\ 
		\nonumber & & \small{ = \left(\begin{array}{ccccc}
		V_{\bar{Q} Q}(r) & \sqrt{ {\J \over 2 \J+1} } V_\text{mix}(r)  & \sqrt{{\J+1 \over 2\J+1}} V_\text{mix}(r) & {1 \over \sqrt{2}} \sqrt{{\J \over 2 \J+1}} V_\text{mix}(r) & {1 \over \sqrt{2}} \sqrt{{\J+1 \over 2\J+1}}V_\text{mix}(r) \\
		\sqrt{{\J \over 2 \J+1}}  V_\text{mix}(r)  &0 & 0 & 0 & 0 \\
		\sqrt{ {\J+1 \over 2\J+1}}  V_\text{mix}(r) & 0  & 0 & 0 & 0 \\
		{1 \over \sqrt{2}} \sqrt{{\J \over 2 \J+1}} V_\text{mix}(r)  & 0 & 0 & 0 & 0 \\
		{1 \over \sqrt{2}} \sqrt{{\J+1 \over 2\J+1}} V_\text{mix}(r) & 0 & 0 & 0 & 0 \\
		\end{array}\right) .} \\
		& &
	\end{eqnarray}
	The confining quarkonium channel is described by $u_{\J}(r)$ with boundary conditions
	\begin{align}
		&u_{\J}(r) \propto r^{\J + 1} \quad &&\text{for} \quad r \rightarrow 0 \\
		&u_{\J}(r) \propto 0 \quad &&\text{for} \quad r \rightarrow \infty,
	\end{align}
	while the incoming wave is a superposition of spherical Bessel functions $j_{L_\text{in}}$. Incoming meson pairs $\bar M M$ or $\bar M_s M_s$ can have either angular momentum $\J-1$ or $\J+1$ (incoming waves with $L_{\text{in}} = \J$ are excluded by parity). Thus, we need to consider four linearly independent superpositions of these incoming waves, defined by $\vec{\alpha} = (\alpha_{\bar{M} M,\J-1}, \alpha_{\bar{M} M,\J+1}, \alpha_{\bar{M}_s M_s,\J-1}, \alpha_{\bar{M}_s M_s,\J+1})$. A simple choice are pure waves, i.e.\ unit vectors for $\vec{\alpha}$. For example $\vec{\alpha} = (1,0,0,0)$ corresponds to a $\bar M M$ wave with $ L_{\text{in}} = \J-1$. The boundary conditions of the emergent waves $\chi_{\bar{M}_{(s)}M_{(s)},L{\text{out}} \rightarrow \J}$ with $L_{\text{out}} \in {\J-1,\J+1}$ define the elements of the $\mbox{T}$ matrix and are given by
	\begin{align}
		& \chi_{\bar{M}_{(s)} M_{(s)},L_{\text{out}} \rightarrow \J} \propto r^{L_{\text{out}}+1} 
		\quad &&\text{for} \quad r \rightarrow 0 \\
		\nonumber & \chi_{\bar{M} M,L_{\text{out}} \rightarrow \J} = i t_{\bar{M}_{(s)} M_{(s)}, L_{\text{in}}; \bar M M, L_{\text{out}}} r h^{(1)}_{L_{\text{out}}}(k r) , && \\
		\label{EQN_t2} & \quad \chi_{\bar{M}_s M_s,L_{\text{out}} \rightarrow \J} =  i t_{\bar{M}_{(s)} M_{(s)}, L_{\text{in}}; \bar M_s M_s, L_{\text{out}}} r h^{(1)}_{L_{\text{out}}}(k_s r) \quad &&\text{for} \quad r \rightarrow \infty ,
	\end{align}
	where $\bar{M} M, L_{\text{in}} = \J-1 \ (\J+1)$ corresponds to $\vec{\alpha} = (1,0,0,0) \ ((0,1,0,0))$ and \\ $\bar{M}_s M_s, L_{\text{in}} = \J-1 \ (\J+1)$ to $\vec{\alpha} = (0,0,1,0) \ ((0,0,0,1))$. The $\mbox{T}$ matrix is
	\begin{align}
		\nonumber & & \hspace{-0.7cm} \mbox{T}_\J = \left(\begin{array}{cccc}
		t_{\bar{M}M, \J-1; \bar{M}M, \J-1} & t_{\bar{M}M, \J+1; \bar{M}M, \J-1}  & t_{\bar{M_s}M_s, \J-1; \bar{M}M, \J-1} & t_{\bar{M_s}M_s, \J+1; \bar{M}M, \J-1} \\
		t_{\bar{M}M, \J-1; \bar{M}M, \J+1} & t_{\bar{M}M, \J+1; \bar{M}M, \J+1}  & t_{\bar{M_s}M_s, \J-1; \bar{M}M, \J+1} & t_{\bar{M_s}M_s, \J+1; \bar{M}M, \J+1} \\
		t_{\bar{M}M, \J-1; \bar{M_s}M_s, \J-1} & t_{\bar{M}M, \J+1; \bar{M_s}M_s, \J-1}  & t_{\bar{M_s}M_s, \J-1; \bar{M_s}M_s, \J-1} & t_{\bar{M_s}M_s, \J+1; \bar{M_s}M_s, \J-1} \\
		t_{\bar{M}M, \J-1; \bar{M_s}M_s, \J+1} & t_{\bar{M}M, \J+1; \bar{M_s}M_s, \J+1}  & t_{\bar{M_s}M_s, \J-1; \bar{M_s}M_s, \J+1} & t_{\bar{M_s}M_s, \J+1; \bar{M_s}M_s, \J+1} \\
		\end{array}\right). \\
		& & \label{eqn:tmatrix}
	\end{align}
	Possibly complex values of the energy $E$, where components of $\mbox{T}_\J$ diverge, are related to masses and widths of both bound states and resonances (see section~\ref{SEC002}).

\section{\label{SEC002}Results}

	To compute the elements of the $\mbox{T}$ matrix \eqref{eqn:tmatrix}, we use two independent methods. The first method reduces the Schr\"odinger equation \eqref{eqn:coupledChannelSE_5x5} by a uniform discretization of the radial coordinate to an ordinary system of linear equations. The second method employs a standard 4th-order Runge Kutta algorithm. The pole positions in the complex energy plane are then determined by a Newton-Raphson algorithm applied to $1/\text{det}(\mbox{T}_\J)$.

	We use the bottom quark mass $m_Q = 4.977 \, \text{GeV}$ from quark models and the spin-averaged masses $m_B = 5.313 \, \text{GeV}$ and $m_{B_s} = 5.403 \, \text{GeV}$ from experiments. $E_{\text{threshold}} = 10.790 \, \text{GeV}$ in the Schr\"odinger equation \eqref{eqn:coupledChannelSE_5x5} is closer to $2 m_{B_s}$ than to $2 m_B$ and reflects that the lattice QCD results from Ref.\ \cite{Bali:2005fu} were obtained with a light $u/d$ quark mass rather close to the physical mass of the $s$ quark. 
	We propagate the uncertainties of the lattice potentials by resampling, i.e.\ we generate 1000 statistically independent samples and repeat all computations on each the 1000 samples. Statistical errors are defined via the 16th and 84th percentile.
	
	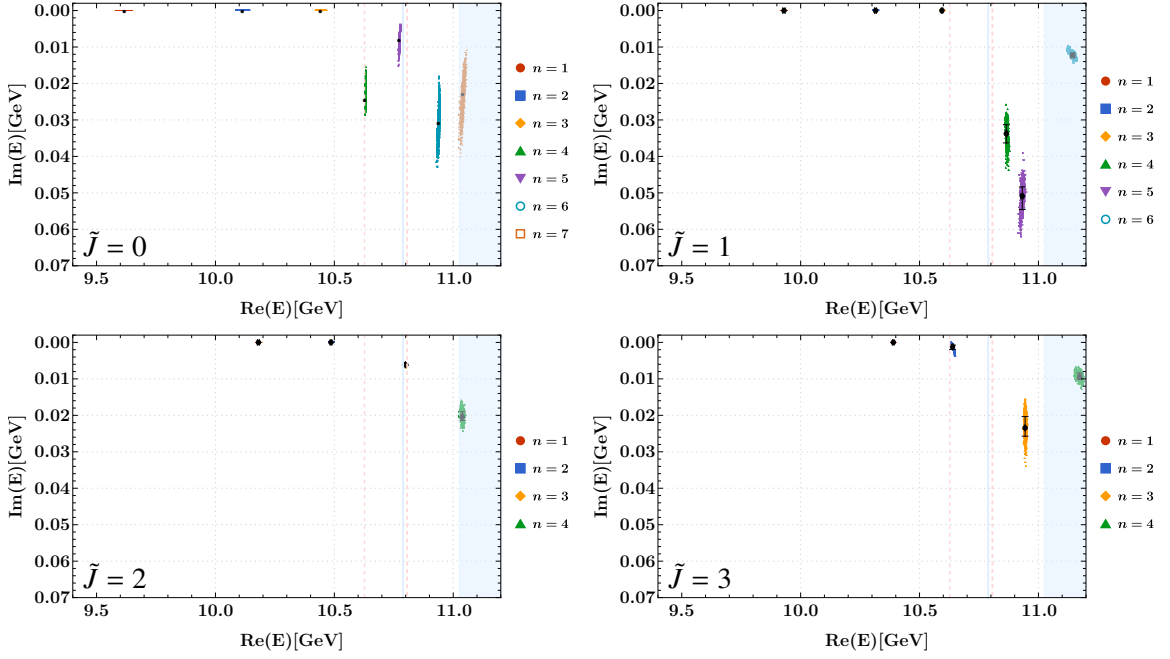
\begin{figure}
	\centering
	\tikzmath{\x = 3.85; \y = 2.2; \imagescale = 0.5; \titlex = 2.3; \titley = 1.1;}
	\begin{tikzpicture}
	\node[inner sep=0] (image1) at (-\x,\y) {\includegraphics[width=\imagescale\textwidth]{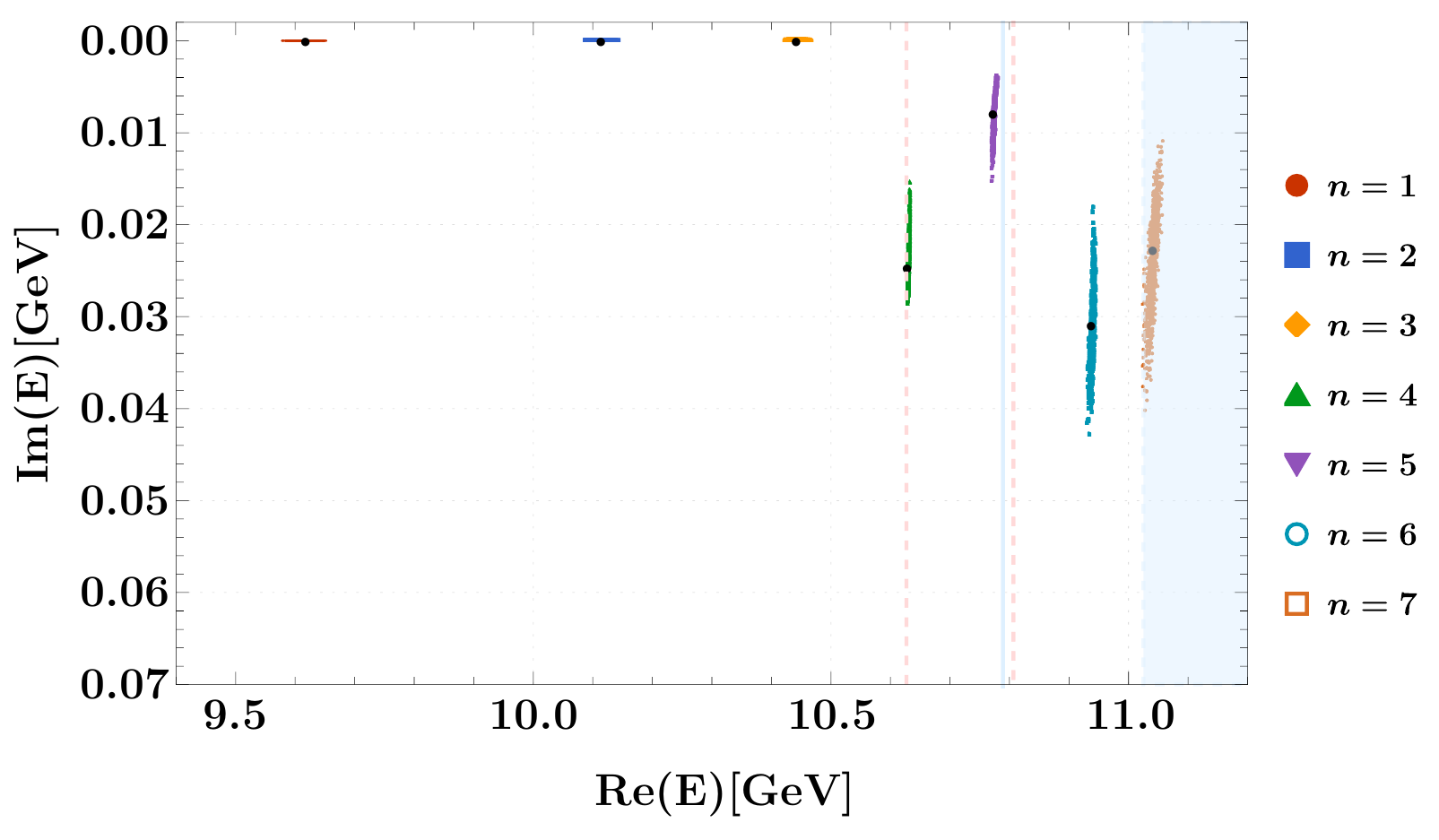}};
	\node at ($(image1)-(\titlex,\titley)$) {$\J = 0$};
	\node[inner sep=0] (image2) at (\x,\y) {\includegraphics[width=\imagescale\textwidth]{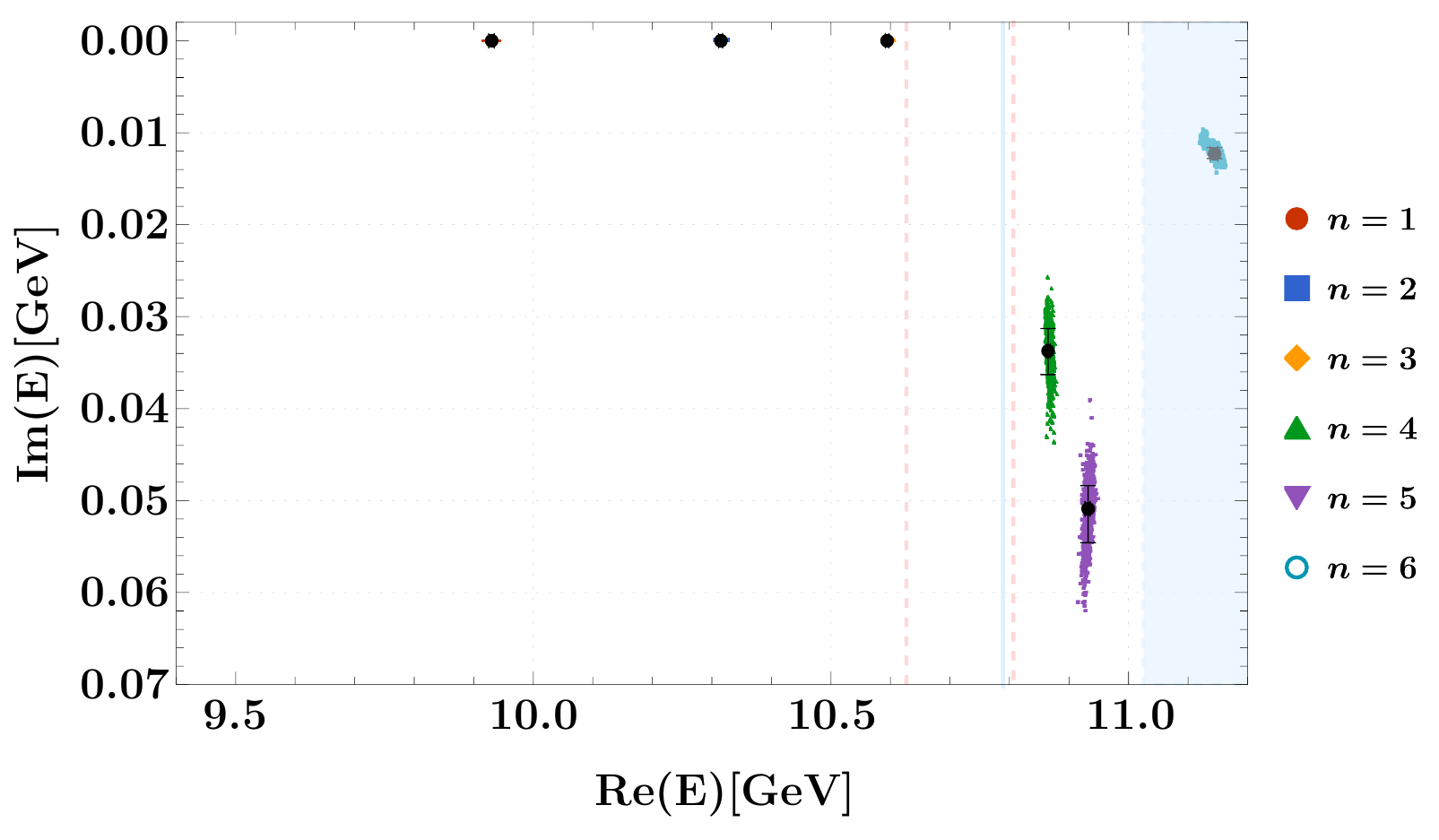}};
	\node at ($(image2)-(\titlex,\titley)$) {$\J = 1$};
	\node[inner sep=0] (image3) at (-\x,-\y) {\includegraphics[width=\imagescale\textwidth]{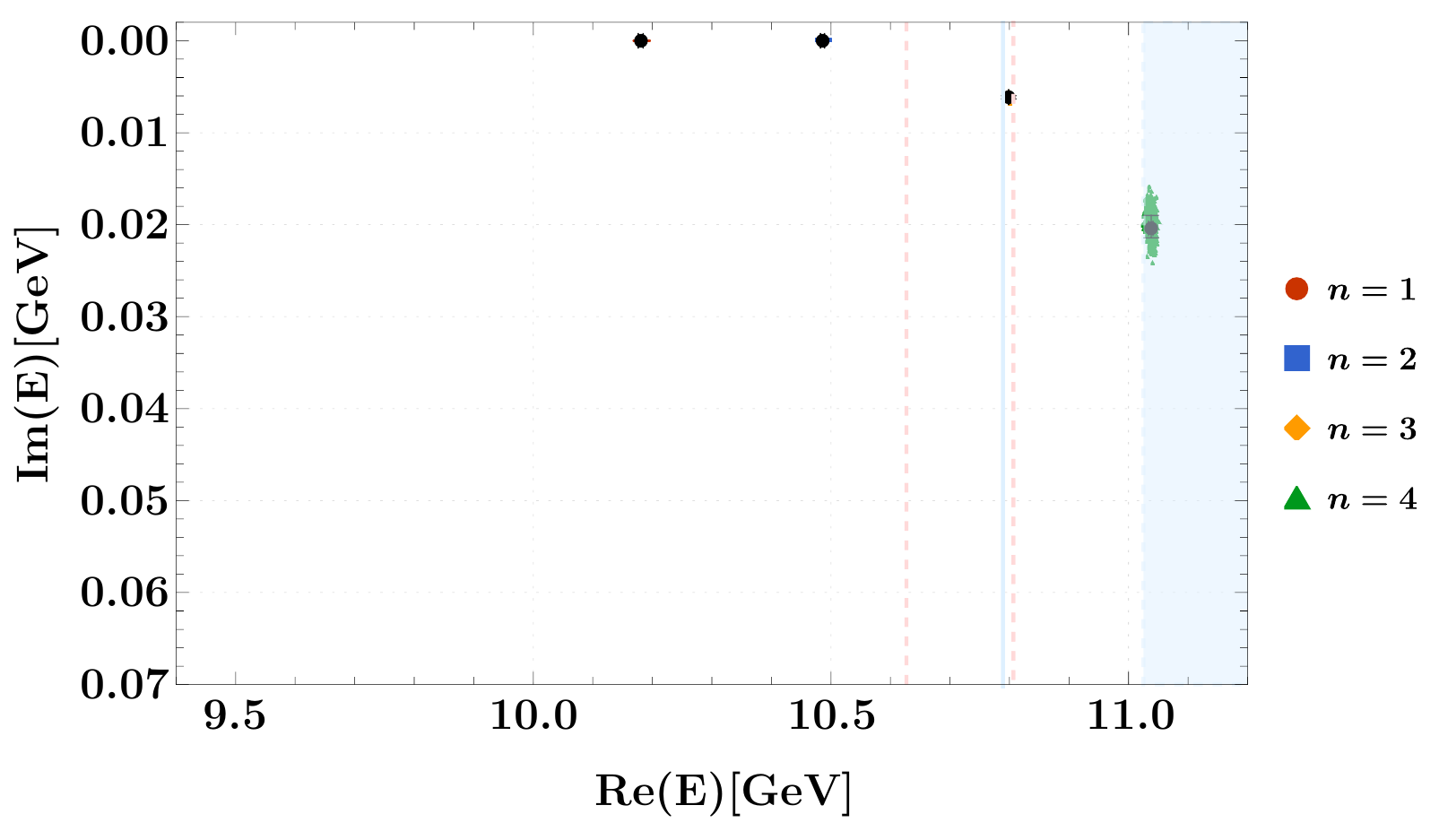}};
	\node at ($(image3)-(\titlex,\titley)$) {$\J = 2$};
	\node[inner sep=0] (image4) at (\x,-\y) {\includegraphics[width=\imagescale\textwidth]{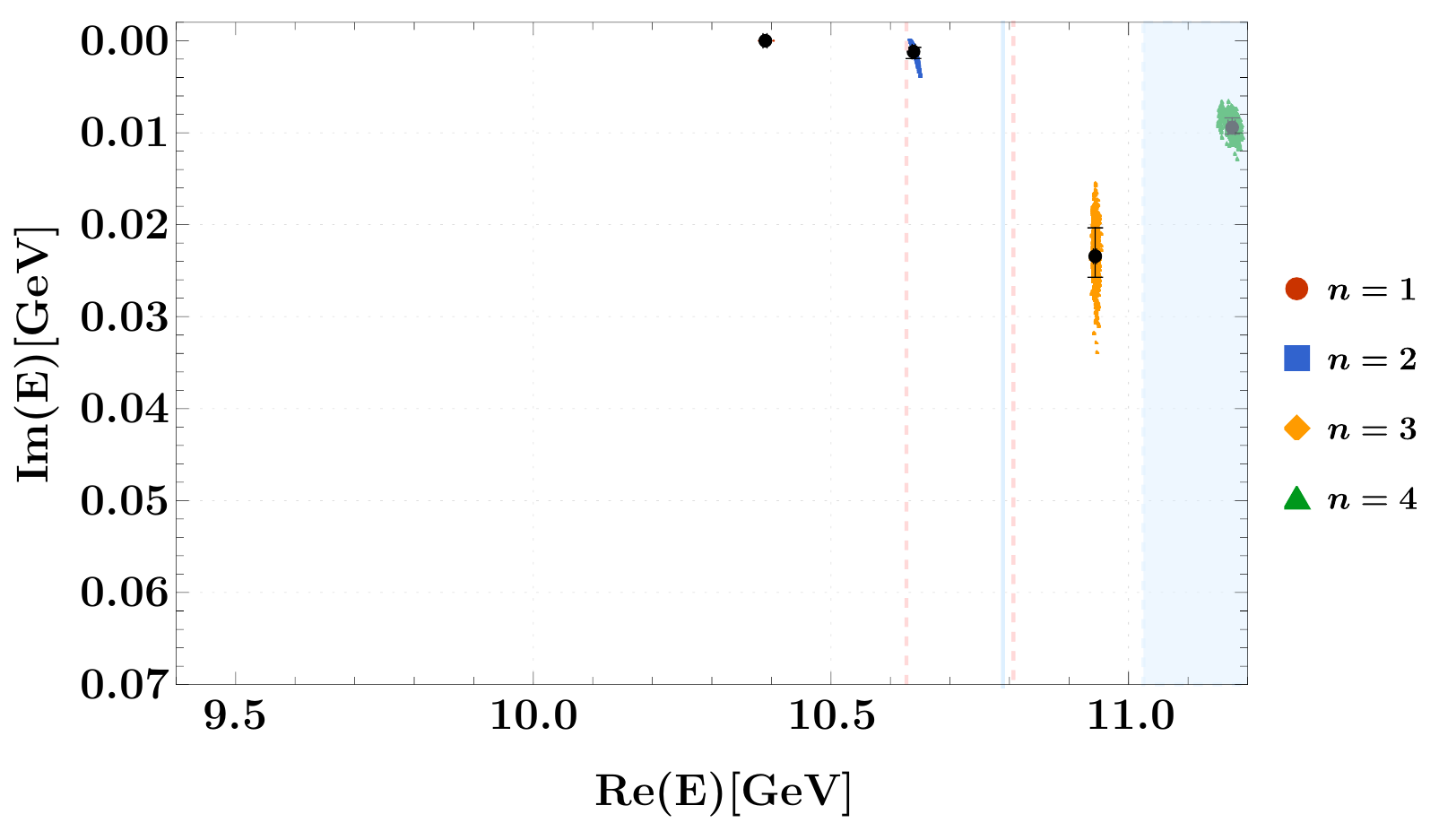}};
	\node at ($(image4)-(\titlex,\titley)$) {$\J = 3$};
	\end{tikzpicture}
	\caption{Positions of the poles of $T_\J$ in the complex energy plane for all bound states and resonances below $11.2 \, \text{GeV}$ for $\J = 0,1,2,3$. Colored point clouds represent results for the 1000 independent samples and black points and bars the corresponding mean values and errors. The vertical dashed lines indicate the spin-averaged $\bar B^{(*)} B^{(*)}$ and $\bar B^{(*)}_s B^{(*)}_s$ thresholds at $10.627 \, \text{GeV}$ and $10.807 \, \text{GeV}$, respectively. The light blue shaded region above $11.025 \, \text{GeV}$ marks the opening of the threshold of one heavy-light meson with negative parity and another one with positive parity. Since this channel is not included in our Schr\"odinger equation, results in this region should not be trusted.}
	\label{fig:polepositions}
	\end{figure}

	In Fig.\ \ref{fig:polepositions} we show all poles of $\mbox{T}_\J$ with corresponding energies below $11.2 \, \text{GeV}$. For each bound state and each resonance there is a differently colored point cloud representing the 1000 samples. Bound states are located on the real axis below the $\bar B^{(*)} B^{(*)}$ threshold at $10.627 \, \text{GeV}$ (indicated by a vertical dashed line), while resonances are above this threshold and have a non-vanishing imaginary part. The complex pole positions $E$ are related to masses and decay widths via $m = \text{Re}(E)$ and $\Gamma = -2 \, \text{Im}(E)$.

\begin{figure*}
		\includegraphics[width=\textwidth]{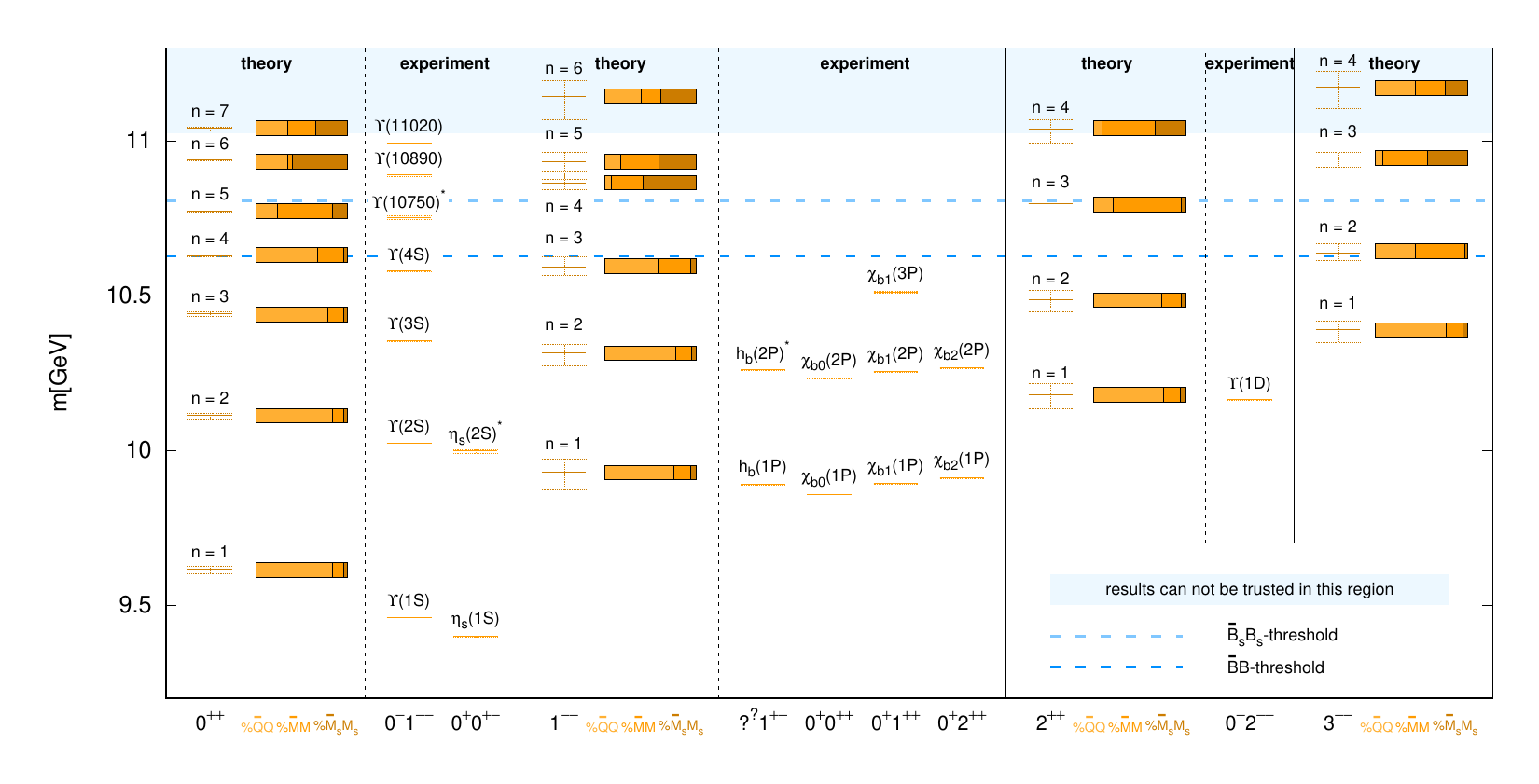}
		\caption{\label{fig:spectrum}Theoretical predictions and experimental results for masses of $I = 0$ bottomonium with \\ $\J^{PC}=0^{++},1^{--},2^{++},3^{--}$. We also show the quarkonium and meson-meson composition defined in Eq.\ \eqref{eqn:percentage}: $\% \bar Q Q$ in light orange, $\% \bar M M$ in medium orange and $\% \bar M_s M_s$ in dark orange. }
	\end{figure*}

	We also determine the quarkonium and meson-meson contributions to the wave function of each state. To this end, we define
	\begin{align}
		\bar Q Q = \frac{Q}{Q + M + M_{s}} \quad , \quad \bar M_{(s)} M_{(s)} = \frac{M_{(s)}}{Q + M + M_{s}} , \label{eqn:percentage}
	\end{align} 
	where 
	\begin{eqnarray}
		& & \hspace{-0.7cm} Q = \int_0^{R_{\text{max}}} \text{d}r \, \Big|u_{\J}(r)\Big|^2 \\
		& & \hspace{-0.7cm} M = \int_0^{R_{\text{max}}} \text{d}r \, \bigg(\Big|\chi_{\bar M M, \J-1 \rightarrow \J}(r)\Big|^2 + \Big|\chi_{\bar M M, \J+1 \rightarrow \J}(r)\Big|^2 \bigg)\\
		& & \hspace{-0.7cm} M_s = \int_0^{R_{\text{max}}} \text{d}r \, \bigg(\Big|\chi_{\bar M_s M_s, \J-1 \rightarrow \J}(r)\Big|^2 + \Big|\chi_{\bar M_s M_s, \J+1 \rightarrow \J}(r)\Big|^2\bigg)
	\end{eqnarray}
	with $R_\text{max} = 2.4 \, \text{fm}$. Results are shown in Fig.\ \ref{fig:spectrum}. A more detailed discussion and plots of the quarkonium and meson-meson percentages as functions of $R_{\text{max}}$ can be found in Ref.\ \cite{Bicudo:2022ihz}.

	In Fig.\ \ref{fig:spectrum} we compare our results with experimentally found bound states and resonances. The pattern of states below or close to the $\bar B^{(*)} B^{(*)}$ threshold is similar to the experimentally observed spectrum:
	\begin{itemize}
		\item $\J = 0$, $n = 1,2,3,4$ states correspond to $\Upsilon(1S)\equiv \eta_s(1S)$, $\Upsilon(2S)$, $\Upsilon(3S)$ and $\Upsilon(4S)$.
		\item $\J = 1$, $n = 1,2,3$ states correspond to $h_b(1P) \equiv \chi_{b0}(1P) \equiv \chi_{b1}(1P) \equiv \chi_{b2}(1P)$, \\ $h_b(2P) \equiv \chi_{b0}(2P) \equiv \chi_{b1}(2P) \equiv \chi_{b2}(2P)$ and $\chi_{b1}(3P)$.
		\item The $\J = 2$, $n = 1$ state corresponds to $\Upsilon(1D)$.
	\end{itemize} 

	The best candidate for the recently found resonance $\Upsilon(10753)$ has $\J = 0$, $n = 5$, is meson dominated and can be classified as a $\Upsilon$ type crypto-exotic state. There is, however, another state very close, which has $\J = 2$, $n = 3$. Moreover, our results support that $\Upsilon(10860)$ corresponds to $\Upsilon(5S)$. For $\Upsilon(11020)$ we again find two candidates, one in an $S$ wave ($\J = 0$, $n = 7$), the other in a $D$ wave ($\J = 2$, $n = 4$). We also find a state close to the $\bar{B}^{(*)}_sB^{(*)}_s$ threshold with a sizable meson-meson component ($\approx 79 \%$), which could have similarities to $X(3872)$ in the charmonium sector. Finally, we predict several states in $P$, $D$ and $F$ waves that have not yet been observed in experiments.

\section*{Acknowledgements}

We acknowledge useful discussions with Gunnar Bali, Marco Cardoso, Eric Braaten, Francesco Knechtli, Vanessa Koch, Sasa Prelovsek and Emilio Ribeiro.

P.B.\ and N.C.\ acknowledge the support of CeFEMA, a research unit funded by FCT under the base and programmatic contract UIDB/04540/2020.
L.M.\ acknowledges support by a Karin and Carlo Giersch Scholarship of the Giersch foundation.
M.W.\ acknowledges support by the Heisenberg Programme of the Deutsche Forschungsgemeinschaft (DFG, German Research Foundation) -- project number 399217702.

Calculations on GPU servers of PtQCD partly supported by NVIDIA were conducted for this research.
Calculations on the GOETHE-HLR and on the FUCHS-CSC high-performance computer of the Frankfurt University were conducted for this research. We would like to thank HPC-Hessen, funded by the State Ministry of Higher Education, Research and the Arts, for programming advice.

\end{document}